\documentclass[a4paper,12pt]{article}

\usepackage[a4paper]{geometry}
\usepackage{amsmath}
\usepackage[final]{graphicx}
\usepackage{bm}
\usepackage{amssymb}

\usepackage{url}

\usepackage[format=plain,indention=1em,margin=10pt,font=normalsize,labelfont=bf]{caption}

\newcommand{\apndxeqn}{ 
\setcounter{equation}{0}%
\renewcommand{\theequation}{\mbox{A\arabic{equation}}} }

\newcommand{\GluonCond}{\big\langle\frac{\alpha_s}{\pi} G^2 \big\rangle}

\newcommand{\bea}{\begin{eqnarray}}
\newcommand{\beq}{\begin{equation}}
\newcommand{\eea}{\end{eqnarray}}
\newcommand{\eeq}{\end{equation}}

\newcommand{\fr}[1]{
             \frac{#1}}

\newcommand{\chibar}{\overline{\chi}}

\newcommand{\ket}{{\rangle}}

\newcommand{\dbr}[1]{\frac{{\rm d}^4 #1}{(2\pi)^4}\:}
\newcommand{\bra}{{\langle}}
\newcommand{\gc}{\bra\fr{\alpha_s}{\pi}G^2\ket}

\newcommand{\qc}{\bra\,\overline{q}q\,\ket}

\usepackage{amssymb}
\newcounter{comment}

{\refstepcounter{comment}%
\begin{quote}
\ttfamily\small$\blacksquare$ \textbf{\underline{Comment} $\sharp$\thecomment:}}%
{\end{quote}}

\begin{document}

\vspace*{4ex}
\begin{center}
\baselineskip=2\baselineskip
\textbf{\LARGE{The Isgur-Wise Function \\ within  a 
  Modified \\ Heavy-Light Chiral Quark Model}}
\\[6ex]
\baselineskip=0.5\baselineskip
{\large Jan~O.~Eeg$^{a,}$\footnote{j.o.eeg@fys.uio.no; corresponding author} and
Kre\v{s}imir~Kumeri\v{c}ki$^{b,}$\footnote{kkumer@phy.hr}} 
\\[4ex]
\begin{flushleft}
\it
$^{a}$Department of Physics, University of Oslo, P.O.B. 1048 Blindern, N-0316 Oslo, 
Norway\\[1.5ex]
$^{b}$Department of Physics, Faculty of Science, University of Zagreb,
 P.O.B. 331, HR-10002 Zagreb, Croatia\\[3ex]
\end{flushleft}
\today \\[5ex]
\end{center}

\begin{abstract}

We consider the Isgur-Wise function $\xi(\omega)$ within a new modified version of a 
heavy-light chiral quark model. While early versions of such models gave 
too small absolute value of the slope, namely $ \xi'(1) \simeq - 0.4$
 to $- 0.3$,
we show how extended version(s) may lead to values around $-1$, in better 
agreement with recent measurements. This is obtained by introducing a
new mass parameter in the heavy quark propagator. 

We also shortly comment on the consequences for the decay modes 
$B~\rightarrow~D\overline{D}$. 

\end{abstract}

\vspace*{2 ex}

\begin{flushleft}
\end{flushleft}

\clearpage

\section{Introduction}

The Isgur-Wise (IW) function $\xi(\omega )$ \cite{iw}, 
the universal function describing a class of
$B(v_b) \rightarrow D(v_c)$ transitions has been studied for many years.
(Here  $v_b$ and $v_c$ are the  four-velocities for the mesons containing a
 $b$,  or a $c$-quark, respectively, and $\omega \equiv v_b \cdot v_c$). 
 While some 
general features of the function follow from heavy quark symmetry \cite{neu},
its more detailed shape has been studied in various model approaches  
\cite{barhi,effr,ebert2,itCQM,AHJOE,ahjoeB,HoSad,CCH,Orsay}.

It is well known that understanding the shape of the IW function, and in
 particular its slope  $\xi'(1)$ at the zero recoil point $\omega =1$ is a
 necessary prerequisite for determination
of the $V_{cb}$ element of the CKM quark mixing matrix. However, there is also another
incentive to study this function in quark models  --- namely, it is 
an essential ingredient in the model description of amplitudes for 
important non-leptonic decays of the type  
$B \rightarrow D \overline{D}, \:  D \, K, \: D \, \pi$. Thus, to approach
such theoretically difficult processes, one first needs
a reasonable description of the IW function.

Some versions of quark models \cite{barhi,effr,AHJOE}
gave a slope $\xi'(1) \simeq - 0.4$ to $- 0.3$,
which is not in agreement with general theoretical expectations expressed
in Bjorken \cite{Bjorken:1990hs} and Uraltsev 
\cite{Uraltsev:2000ce} sum rules which together imply 
$ - \xi'(1) \ge 3/4 $.
Also, a combined fit \cite{Barberio:2008fa} to results of experimental 
measurements of $B\to D^* l \nu$ decays gives
$- \xi'(1) = 1.16 \pm 0.05$, and, although dispersion
of experimental results is large leading to a small confidence level
of this fit ($\approx 1 \%$), it seems reasonable to assume that absolute value
of the slope cannot be significantly smaller than~\,1.
In this paper we propose a modified version of the model in \cite{AHJOE},
which has a particular feature of explicit inclusion of the
gluon condensate effects, enabling consistent estimation of non-factorizable
amplitudes \cite{BEF,EHP,EFZ,EFP,MacDJoe}, and we demonstrate that
this model gives a  satisfactory description of the IW function slope.

There are two slightly different philosophies within the 
family of heavy-light chiral
quark models. In Refs. \cite{barhi,effr,AHJOE} the focus is on the 
{\em bosonization} procedure. The quark Lagrangian is bosonized by attaching 
meson fields to quark loops at zero external momentum,
 thereby integrating out the quarks.
External momenta then correspond to derivatives of meson fields.
 On the other hand,
in the approach of the Bari group \cite{itCQM} the external momenta are kept
in the loop integral and mass differences between heavy mesons and the heavy 
quark appear in the final result. 
This mostly works 
fine in \cite{itCQM}, but a problem seems to arise when one tries to describe
transitions between heavy mesons with different masses. 

In this paper we will work with zero external momenta (focus on bosonization),
but  introduce a parameter $\Delta $ in the heavy
quark propagator, which will, for positive values of $\Delta$, correspond
to an extra dynamical mass of the heavy quark. 

In Sect.~\ref{sect:hlcqm} we describe this extended version of heavy-light
chiral quark model. In Sect.~\ref{sect:bosonisation} we show how the model
is bosonised and how the model parameters are related to physical quantities.
In Sect.~\ref{sect:iw} we calculate the IW function, and then we conclude.
Appendix~\ref{app:loop} contains recursion formulae and expressions for the
relevant heavy-light loop integrals.

\section{Heavy-Light Chiral Quark Model (HL$\chi$QM)}
\label{sect:hlcqm}

The total Lagrangian describing both quark
 and meson fields is \cite{AHJOE}:
\begin{equation}
{\cal L} =  {\cal L}_{HQET} +  {\cal L}_{\chi QM}  +   {\cal L}_{Int} \; ,
\label{totlag}
\end{equation}
where \cite{neu}
\begin{equation}
{\cal L}_{HQET} =  \overline{Q_{v}} \, \left( i v \cdot D \, 
- \, \Delta  \right) Q_{v} 
+ {\cal O}(m_Q^{- 1})
\label{LHQET}
\end{equation}
is the Lagrangian for Heavy Quark Effective Field Theory (HQEFT), with 
the mentioned extra mass added. 
The heavy quark field  $Q_v$
annihilates  a heavy quark  with velocity $v$ and mass
$m_Q$. Moreover,  
$D_\mu$ is the covariant derivative containing the gluon field
(eventually also the photon field).
In \cite{AHJOE}, the  ${\cal O}(m_Q^{- 1})$ term was also considered, but it
 will not be needed in this paper, because $1/m_c$ and $1/m_b$ corrections 
will not be calculated.

The light quark sector is described by the Chiral Quark Model ($\chi$QM),
having a standard QCD term and a term describing interactions between
quarks and  pseudo-scalar light mesons: 
\begin{equation}
{\cal L}_{\chi QM} =   \bar{q}(i\gamma^\mu D_\mu  -  {\cal M}_q) q
  -     m(\bar{q}_R \Sigma^{\dagger} q_L   +    \bar{q}_L \Sigma q_R) \; \, , 
\label{chqmU}
\end{equation}
where $q^T  =  (u,d,s)$ is the light quark field triplet. The left- and
 right-handed
 projections $q_L$ and $q_R$ are transforming after $SU(3)_L$ and $SU(3)_R$,
respectively. ${\cal M}_q = {\rm diag}(m_u, m_d, m_s)$ is the current quark
 mass matrix,
$m$ is the ($SU(3)$ invariant) dynamical  mass of light quarks, and 
$\Sigma  =   \exp(2 i \Pi/f_\pi)$, where $\Pi$ is a 3 by 3 matrix containing
the pseudo-scalar meson octet ($\pi, K, \eta$) in a standard way.

 There is also a ``rotated  version'' of  the $\chi$QM
with  flavour-rotated quark fields $\chi$ given by:
\begin{equation}
\chi_L  =   \xi^\dagger q_L \quad ; \qquad \chi_R  =   \xi q_R \quad ; \qquad 
\xi \cdot \xi  =   \Sigma \; .
\label{rot}
\end{equation}
(This field $\xi$ containing the light mesons should be distinguished from the
IW function $\xi(\omega)$.)  
In the rotated version, the chiral interactions are transformed into the
kinetic term,  while the interaction term proportional to $m$ in (\ref{chqmU}) 
becomes a pure (constituent) mass term \cite{chiqm,BEF}:
\begin{equation}
{\cal L}_{\chi QM} =  
\chibar \left[\gamma^\mu (i D_\mu   +    {\cal V}_{\mu}  +  
\gamma_5  {\cal A}_{\mu})    -    m \right]\chi 
  -     \chibar \widetilde{M_q} \chi \;  , 
\label{chqmR}
\end{equation}
where the vector and axial vector fields 
${\cal V}_{\mu}$ and  
${\cal A}_\mu$ are given by:
\begin{equation}
\left. \begin{aligned}{\cal V}_{\mu} \\ {\cal A}_\mu \end{aligned} \right\} 
\, = \, 
\pm \fr{i}{2}(\xi^\dagger\partial_\mu\xi
 \pm \xi\partial_\mu\xi^\dagger ) \; \, , 
\nonumber
\end{equation}
and
\bea
\widetilde{M_q} \equiv
\xi^\dagger {\cal M}_q^\dagger\xi \, L  \; 
+ \;  \xi {\cal M}_q ^\dagger \xi \; R \; \, .
\eea
Here $L$ is the left-handed projector in Dirac space, 
$L =  (1 -  \gamma_5)/2$,
and $R$ is the corresponding right-handed projector.
The Lagrangian (\ref{chqmR}) is manifestly invariant under the unbroken
$SU(3)_V$ symmetry.
In the light sector, the various pieces of the Lagrangian
describing strong interactions of mesons
can be obtained by integrating out the constituent quark fields $\chi$,
and these pieces can be written in terms of the  fields ${\cal A}_\mu$
and $\widetilde{M}_q$ 
which are manifestly invariant
under local $SU(3)_V$ transformations.

In the heavy-light case, the generalization of the
meson-quark interactions of the pure light sector  $\chi$QM
is given by the following $SU(3)$ invariant Lagrangian:
\begin{equation}
{\cal L}_{Int}  =   
 -   G_H \, \left[ \chibar_a \, \overline{H_{v}^{a}} \, Q_{v} \,
  +     \overline{Q_{v}} \, H_{v}^{a} \, \chi_a \right]   +   
 \fr{1}{2G_3}Tr\left[ \overline{H_{v}^{a}}\,  H_{v}^{a}\right] \; ,
\label{Int}
\end{equation}
where $G_H$ and $G_3$ are  coupling constants, and
 $H_{v}^{a}$ is the heavy meson field  containing
 a spin zero and spin one boson:
\begin{eqnarray}
&H_{v}^{a} & \equiv  P_{+} (P_{\mu}^{a} \gamma^\mu -     
i P_{5}^{a} \gamma_5)\; , \nonumber \\
&\overline{H_{v}^a}
& =  \gamma^0 (H_{v}^a)^\dagger \gamma^0
 =  \left[(P_{\mu}^{a})^{\dagger} \gamma^\mu 
 -   i (P_{5}^{a})^\dagger \gamma_5\right] P_{+} \; , \label{barH}
\end{eqnarray}
where
\begin{equation}
P_{\pm} \equiv  \frac{1 \pm \gamma \cdot v}{2} \; .
 \qquad H_v \gamma \cdot v \,= - H_v  \; , \quad
 \gamma \cdot v \, \overline{H_v}= - \overline{H_v}  \;.
\label{proj}
\end{equation}
The field $P^{a}_{5} (P^{a}_{\mu})$ annihilates a heavy-light meson,
 $0^{-}(1^-)$,  with velocity $v$.
The index $a$ runs over the light quark flavours $u, d, s$, and
the projection operators $P_{\pm}$ have the  property
\begin{equation}
P_{\pm}\gamma^\mu P_{\pm}  =  \pm\, P_{\pm}\, v^\mu\,P_\pm \; .  
\label{projrel}
\end{equation}
Note that in Refs. \cite{barhi,effr,itCQM,ebert2} $G_H=1$ is used. However,
 in that case
one uses a renormalization factor  for the heavy meson fields $H_v$,
which is equivalent to the approach in \cite{AHJOE} and here.
The term $\propto 1/G_3$ is (partially) cancelled by a self-energy
 loop of order $G_H^2$, and determines the mass difference between the
 heavy quark and the corresponding heavy meson. 

\begin{figure}[tb]
\centerline{\includegraphics[scale=1.2]{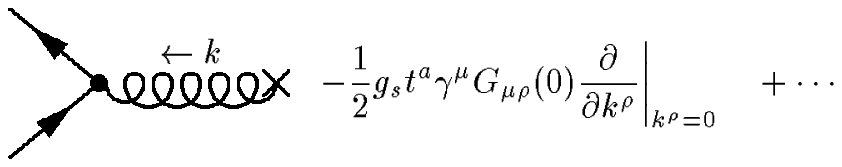}}
\caption{Feynman rule for the light quark - soft gluon vertex.}
\label{fig:gc}
\end{figure} 

 In our model, the hard gluons are considered to be integrated out and we are
left with soft gluonic degrees of freedom. These soft gluons can be
described using the external field technique, and their
effect will be parameterized by vacuum expectation values, 
such as the gluon condensate $\gc$. Gluon condensates with higher 
dimensions could also be included, but we truncate the expansion by keeping
 only the one with lowest dimension.
When calculating the soft gluon effects in terms of the gluon condensate,
we follow the prescription given in \cite{nov}.
The calculation  is easily carried out in the
Fock-Schwinger gauge, where one can expand the gluon field as
\begin{equation}
A_\mu^a(k) =   -  \fr{i(2\pi)^4}{2}G_{\rho\mu}(0)\fr{\partial}{\partial
k_\rho}\delta^{(4)}(k)  +  \cdots\, .
\label{eq:Afield}
\end{equation}
Since each vertex in a Feynman diagram is accompanied by an
integration, we get the Feynman rule given in figure \ref{fig:gc}.
The gluon condensate is obtained by averaging in colour space which yields
the following replacement rule:
\begin{equation}
g_s^2 G_{\mu \nu}^a G_{\alpha \beta}^b  \; \rightarrow \fr{4 \pi^2}{(N_c^2-1)}
\delta^{a b} \gc \frac{1}{12} (g_{\mu \alpha} g_{\nu \beta} -  
g_{\mu \beta} g_{\nu \alpha} ) \; .
\end{equation}

\section{Bosonization within the HL$\chi$QM}\label{sec:strong}
\label{sect:bosonisation}
 
The interaction term ${\cal{L}}_{Int}$ in (\ref{Int}) can now 
be used to bosonize the model, i.e. to
integrate out the quark fields. This can be done in the path 
integral formalism and the result is formally a functional determinant. This
determinant can be expanded in terms of Feynman diagrams, by attaching the
 external fields $H_v^{a}, \overline{H_v^{a}},  {\cal{V}}^\mu,
  {\cal{A}}^\mu$ and $\widetilde{M}_q$ of section \ref{sect:hlcqm} to 
quark loops.
Some of the loop integrals will be divergent and, analogously to the pure
light sector case \cite{chiqm,pider,epb,BEF}, they have to be related to
 physical parameters.
The resulting strong chiral Lagrangian 
has the following form \cite{wise1,itchpt,wiseR,Grin,GriBo,IWSte,EJenk}: 
\bea
{\cal L}_{Str}\, = 
  \, - \, 
Tr\left[\overline{H_{a}}(iv\cdot {\cal D})H_{a}\right] 
\,  - \, g_{\cal A} \,                
Tr\left[\overline{H_{a}}H_{b}\gamma_\mu\gamma_5 {\cal
A}^\mu_{ba}\right] \; + \; .... \; ,
 \label{LS1}
\eea
where the dots indicate other terms of higher order in the chiral expansion,
 and the covariant derivative ${\cal D} $ contains both the photon field
 and the field ${\cal V}$.
The $1/m_Q$ suppressed terms have been discarded in the present paper.

\begin{figure}[tb]
\centerline{\includegraphics[scale=1.0]{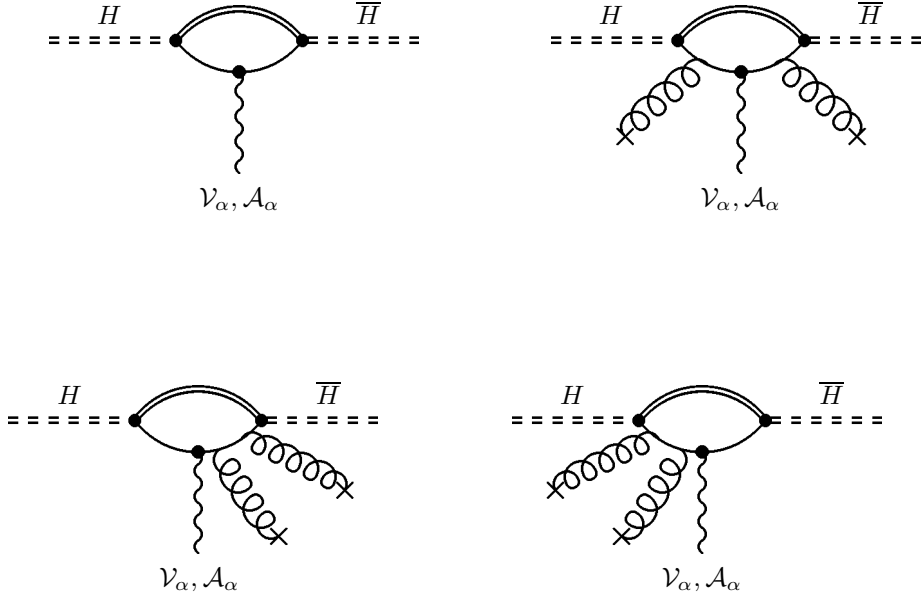}}
\caption{Coupling to vector and axial vector current}
\label{fig:va}
\end{figure}

The Feynman diagrams 
responsible for the kinetic  and axial vector terms in (\ref{LS1}) are shown in 
Fig. \ref{fig:va}. As mentioned in the Introduction,
these two diagrams are calculated at zero external heavy meson momentum.
The non-gluonic loop integral (first on Fig. \ref{fig:va}) 
for the strong vector or axial vector current is
\bea
J^\mu_X \,  = \, - N_c \int 
\dbr{k}
 {\rm Tr}\Big\{(-i G_H \overline{H_v}) \, i S_v(k)
(-i G_H H_v) \,  i S(k) \Gamma^\mu_X  \, i S(k) \Big\} \;,
\label{CurrX}
\eea
where
 $\Gamma^\mu_{{\cal V}} = \gamma^\mu$ and  
$\Gamma^\mu_{{\cal A}} = \gamma^\mu \, \gamma_5$ for couplings to 
$X \, = \, {\cal V}$ and $X \, = \, {\cal A}$. Further, 
$S_v(k)$ and  $S(k)$ are the heavy quark propagator and 
the standard light quark propagator, respectively:
\begin{equation}
S_v(k) \; = \;  \frac{ P_+(v)}{(v  \cdot k  - \Delta)}  
\qquad ; \quad  S(k) \; = \; 
 \frac{\gamma \cdot k \, + \, m}{( k^2 \, + m^2)} \; \, . 
\end{equation}
In previous papers \cite{barhi,effr,AHJOE,itCQM},  $\Delta=0$ was assumed, 
but here we let $\Delta \neq 0$.

The gluonic part of the bosonized currents  for one of the diagrams
(lower left on Fig.~\ref{fig:va}) is
\bea
J^\mu_{X,G1} \,  = \, -  \int 
\dbr{k}
 {\rm Tr} \Big\{(-i G_H \overline{H_v}) \, i S_v(k)
(-i G_H H_v)  \, i S(k) \Gamma^\mu_X  \, i S(k)   \nonumber \\
 \left[i g_s \gamma^\alpha A_\alpha(q_2) \right] \, i S(k-q_2) 
\left[i g_s \gamma^\alpha A_\alpha(q_1) \right] \, i S(k-q_1 - q_2)
 \Big\} \; \, ,
\label{CurrXG1}
\eea
where gluon fields are represented by the expression 
from Eq. (\ref{eq:Afield}), 
and it is understood that the derivatives with respect to soft gluon 
momenta are to be applied to the whole integrand.
There are two more diagrams with different ordering
of gluon and (axial) vector vertices and after
 adding all four
diagrams from Fig.~\ref{fig:va} we obtain
\bea
J^\mu_{X,{\rm Tot}} \,  = \,   - \, g_X \, 
Tr \left\{\overline{H_v} \, H_v \,  \Gamma^\mu_X 
\right\} \; \, ,
\label{CurTot}
\eea
where 
\bea
g_X \,  = \,  i G_H^2 \, N_c\left\{ R_X  \, - \, 
\frac{\pi^2}{24 N_c}\, Z_X \langle \frac{\alpha_s}{\pi} G^2 \rangle
\right\}  \; \, ,
\label{KX}
\eea

\begin{align}
 R_{{\cal V}} & = -2(m-\Delta)I_{2}-I_{1,1}-
2\Delta(m-\Delta) I_{2,1} \; \, , \\[1ex]
 Z_{{\cal V}} & = 144 m I_4 +192 m^2 (m-\Delta)I_5 
+ 24 m ( m + 6\Delta)I_{4,1} \nonumber \\[1ex]
&\quad +192 m^2 \Delta (m - \Delta)I_{5,1} \; \, ,
\end{align}
and for the axial case
\begin{align}
R_{{\cal A}} & = -\frac{2}{3}(3m-\Delta)I_{2}+\frac{1}{3}I_{1,1}+
\frac{2}{3}(m-\Delta)(2m-\Delta) I_{2,1}  \; \, , \\[1ex]
Z_{{\cal A}} & =  48 m I_4 +64 m^2 ( 3m-\Delta)I_5 - 
  8 m (13 m - 6\Delta)I_{4,1} \nonumber \\[1ex]
&\quad -64 m^2 (m- \Delta)(2m-\Delta)I_{5,1} \; \, . 
\end{align}
The loop integrals $I_{n}$ and $I_{n,r}$ occurring in the expressions above
are defined in Appendix~\ref{app:loop}, where
also expressions for the finite ones are given.
The integrals $I_2$, $I_{1,1}$ above and $I_1$ from
Eq. (\ref{eq:alphaH}) below are
logarithmically, linearly, and quadratically divergent,
respectively. They will be expressed in terms of model parameters.
The negative parity axial coupling constant $g_{{\cal A}}$ is taken as model
input parameter, $g_{{\cal A}} = 0.59$, and $g_{{\cal V}} = 1$
 (i.e. the normalization).

Within the pure light sector the logarithmically and quadratically divergent
integrals are related to the pion decay
constant $f_\pi$ and the quark condensate $\qc$ in the following way
\cite{pider,epb,BEF}:
\begin{equation}\label{I2}
f_\pi^2 =    -  i4m^2N_cI_2 +  \fr{1}{24m^2}\gc \; ,
\end{equation}
\begin{equation}\label{I1}
\qc = -4imN_cI_1-\fr{1}{12m}\gc \; .
\end{equation}
This is obtained by relating loop diagrams to physical quantities
analogously to Eqs. (\ref{norm}) and (\ref{ga}) below. (Here the a priori 
divergent integrals $I_{1}$ and $I_2$ have to be interpreted as 
the regularized ones.) 
Since the pure light sector is a part of our model, we keep these 
relations in the heavy-light case studied here.
In addition, in
the heavy-light sector the (formally) linearly divergent integral $I_{1,1}$ 
will also appear. 
It will be related to the physical value of $g_{\cal A}$ 
using Eq. (\ref{KX}) for $X={\cal A}$.

Eliminating thus $I_{1,1}$ from the (\ref{KX})
and inserting the expression for $I_2$ obtained from (\ref{I2}) 
we find the following expression for $G_H$:  
\begin{equation}
G_H^2  =   \fr{2m}{f_\pi^2} \, \rho_\Delta \; \, , 
\label{GHrho}
\end{equation}
where the quantity $\rho_\Delta$ is of order one and  given by
\begin{equation}
\label{rhoDel}
\rho_\Delta \, \equiv \, \fr{1 +  3g_{\cal A}}{4 \left(1 -
 \frac{\Delta}{2 m} + 
 \frac{N_c m^2}{8 \pi f_\pi^2} \kappa_\Delta 
-\frac{\eta_{1}^{\Delta}}{m^2}\gc \right) }
\;,
\end{equation}
where 
\begin{equation}
\kappa_\Delta  =  i 16 \pi  m (1 - \frac{\Delta}{m})^2 I_{2,1} \;, 
\label{kappa}
\end{equation}
and 
\begin{equation}
\eta_1^\Delta  =    \frac{1}{12} \left[1 - \frac{\Delta}{2 m} + 
i \frac{\pi^2 m^3}{2} (Z_{{\cal V}} + 3 Z_{{\cal A}} ) \right]\;.
\label{eta1}
\end{equation}

Let us now consider these relations in two characteristic limiting
cases: $\Delta \to 0$ and $\Delta \to m$.

In the limit $\Delta \rightarrow 0$, (\ref{KX}) reduces to
\begin{align}
 1& = -  iG_H^2N_c \, \Big\{I_{3/2}  +   2mI_2  +  
\fr{i \kappa_{{\cal V}}}{N_c \, m^3}\gc \Big\} \;, \label{norm} \\
 g_{\cal A}& = i G_H^2 N_c \bigg\{\fr{1}{3}I_{3/2} -   2mI_2  - 
i   \fr{m}{12 \pi} -   \fr{i \kappa_{{\cal A}}}{N_c \, m^3}\gc\bigg\} \;,
\label{ga}
\end{align}
where
\begin{equation}
I_{3/2}\equiv (I_{1,1})_{\Delta\to 0} \; ; \quad 
\kappa_{{\cal V}} \,  = \, - \, \kappa_{{\cal A}} \; = \; \frac{(8 - 3 \pi)}{384}    
 \; .
\label{kappas}
\end{equation}
This is the result of \cite{AHJOE}, except for the sign of $\kappa_{{\cal A}}$
which is wrong there. Numerically, this change has no dramatic consequences.
In the limit  $\Delta \rightarrow 0$ we also have
\begin{equation}
\kappa_\Delta \rightarrow 1 \; ; \; \,  \eta_1^\Delta \, \rightarrow \,
 \eta_1 \, = \, \frac{1}{12} -  (\kappa_{{\cal V}} + 3 \kappa_{{\cal A}}) \,
 = \frac{8-\pi}{64} \;.
\label{etah}
\end{equation}
Eliminating $I_2$ we obtain the relation for $I_{3/2}$:
\begin{equation}
-i N_c I_{3/2} \; = \; \frac{3(1-g_{{\cal A}})}{4 G_H^2} \, - \,
 \fr{m}{16 \pi} -   \fr{3 (\kappa_{{\cal A}} -\kappa_{{V}})}{4 m^3}\gc
 \;,
\label{I32}
\end{equation}
which will replace the Eq. (39) of \cite{AHJOE}.

In the limit $\Delta \rightarrow  m$, which means that heavy and light
 quarks have the same {\em constituent} mass, we obtain
\begin{align}
 1& = G_H^2  \, \Big\{ - i N_cI_{11}  \, + \,    
\fr{1}{240 m^3}\gc \Big\} \;, \label{normD} \\
 g_{\cal A}& =  G_H^2 N_c \bigg\{ \fr{1}{3} i N_c I_{11} -  
 \frac{4}{3} m I_2  - 
   \fr{3}{240 m^3}\gc\bigg\} \;.
\label{gaD}
\end{align}
Then, for  $\Delta \rightarrow  m$, there is a simplification 
because $\kappa_\Delta \rightarrow 0$ and $\eta_1^\Delta \rightarrow 0$
and thereby
\begin{equation}
\label{rhom}
\rho_\Delta \, \rightarrow  \, \frac{1}{2} \, (1 +  3g_{\cal A}) \;.
\end{equation}

\begin{figure}[tb]
\begin{center}
\includegraphics[scale=0.7]{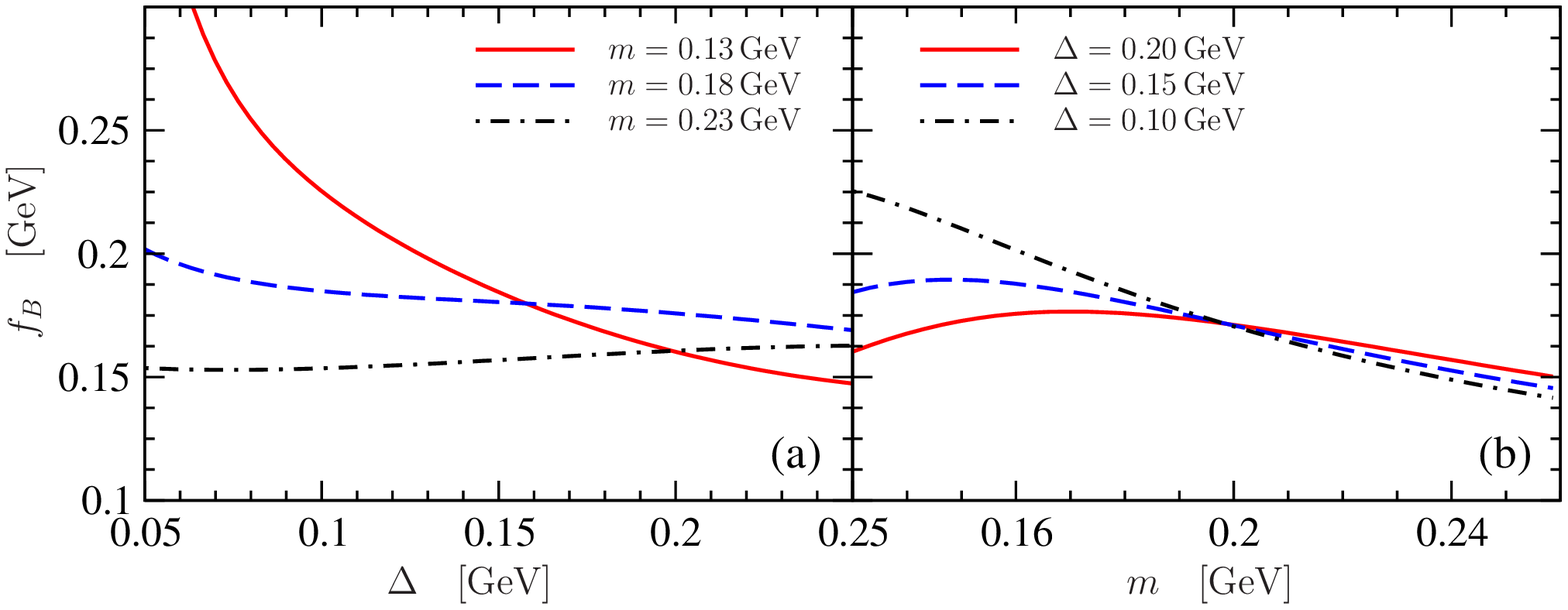}
\end{center}
\caption{The decay constant $f_B$ in dependence on $m$ and $\Delta$.
The  condensates are taken to be
$\qc = (-0.27\,{\rm GeV})^3$ and $\gc = (0.32\,{\rm GeV})^4$.
}
\label{fig:fB}
\end{figure}

The gluon condensate may be related to the
 matrix element of the chromomagnetic interaction \cite{neu}: 
\begin{equation}
3 \lambda_2 \, = \,  \mu_G^2(H) =  \frac{C_{\text{M}}(\mu)}{2M_H}
\bra H|\bar{Q_v}\frac{1}{2}\sigma\cdot GQ_v|H\ket\, =  
 \, \frac{3}{2} m_Q (M_{H^*} -  M_H)
 \; .  
\label{gc2}
\end{equation} 
Such a link was used in \cite{AHJOE}, but we will not use it because it
formally belongs to $1/m_Q$ corrections, which are not considered here. 
Also, it turns out that such a choice doesn't lead to any
dramatic differences, when numerical results in $\Delta \to 0$ limit are 
compared to those from \cite{AHJOE}.

Within the full theory (Standard Model) at quark level, the weak current is
\begin{equation}
J_f^\alpha  =  \overline{q}_f \gamma^\alpha (1 -  \gamma_5) Q \;,
\label{Lcur}
\end{equation}
where $Q$ is the heavy quark field in the full theory.
Within HQEFT this current will,  below the renormalization scale
 $\mu  =   m_Q \, (=m_b, m_c)$, be modified in the following way \cite{neu}:
\begin{equation}
J_f^\alpha
 =  {\chibar}_a \xi^{\dagger}_{af} \Gamma^\alpha  Q_v     
\, + {\cal O}(m_Q^{-1}) \; ,
\label{modcur}
\end{equation}
where 
\begin{equation}
 \Gamma^\alpha \; = \; C_\gamma \, \gamma^\alpha \, L \; 
+ \, C_v \,v^\alpha \, R \; .
\label{modcurgam}
\end{equation}

\begin{figure}[tb]
\centerline{\includegraphics[scale=1.2]{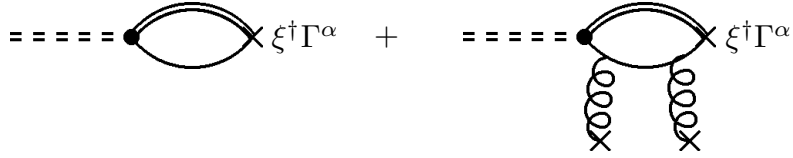}}
\caption{Diagrams for bosonization of the left handed quark current to leading order, determining $\alpha_H$}
\label{fig:f+}
\end{figure}

\begin{figure}[tb]
\begin{center}
\includegraphics[scale=0.62]{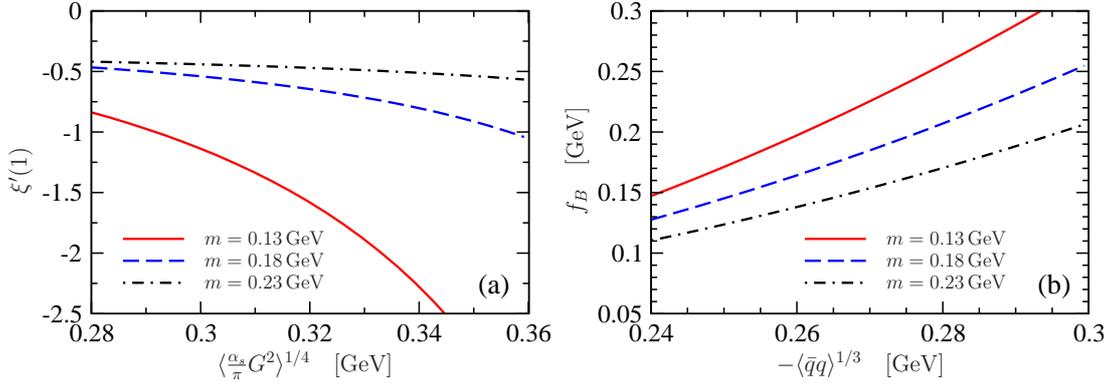}
\end{center}
\caption{The slope of Isgur-Wise function $\xi'(1)$ in dependence on the 
gluon condensate for various choices of the light quark constituent mass $m$ (a),
and the decay constant $f_B$ in dependence on the quark condensate (b), for
$\Delta = 0.1\, {\rm GeV}$.
Complementary plots are not so interesting because the
dependence of $\xi'(1)$ on the \emph{quark} condensate and the 
dependence of $f_B$ on the \emph{gluon} condensate are rather small.
}
\label{fig:cond}
\end{figure}

Bosonising this current by calculating diagrams from Fig.~\ref{fig:f+}
we obtain to zeroth order in the axial
field and to first order in the gluon condensate the weak current
\begin{equation}
J_f^\alpha    =    \frac{\alpha_H}{2} Tr\left[\xi^{\dagger}_{hf}\Gamma^\alpha
  H_{vh}\right]
  \; ,\label{J(0)}
\end{equation}
where heavy meson decay constant is given by
\begin{equation}
f_{H} = \frac{C_\gamma + C_v}{\sqrt{M_H}} \alpha_H \;,
\end{equation}
with $C_{\gamma}=1.077$ and $C_{v}=0.0489$ being Wilson coefficients
 within HQEFT \cite{neu}.
In addition $f_B$ and  $f_D$ have chiral corrections of 
order 20 MeV (see \cite{AHJOE} and references therein),
 as well as  $1/m_Q$ corrections not considered here.
From the diagrams in Fig. \ref{fig:f+},  we obtain:
\begin{align}
 \alpha_{H} & = -2 i  G_H 
N_c \Big\{ -I_1 + ( m-\Delta)I_{1,1} \nonumber \\
& \quad + \frac{m \pi^2}{N_c} \GluonCond \big[-m I_4 + I_{3,1}
+ m ( m-\Delta) I_{4,1} \big] \Big\} \;,
\label{eq:alphaH}
\end{align}
where divergent integrals $I_1$ and $I_{1,1}$ will be expressed
in terms of model parameters. Since this is the only predicted
quantity depending on $I_1$ it is easy to accommodate 
physical values of $f_H$, using (\ref{I1}) 
and adjusting the quark condensate, see Fig.~\ref{fig:cond}(b).
In the limit $\Delta \rightarrow 0$
\begin{equation}
\alpha_H \equiv  -  2iG_HN_c\left(  -  I_1  +  mI_{3/2} +
 \fr{i \, \kappa_H}{N_c \, m^2}\gc\right) \; .
\end{equation}
where $\kappa_H \, = \, (3 \pi -4)/384$, in agreement with \cite{AHJOE}.
Furthermore, eliminating divergent integrals, one obtains
\begin{equation}
\alpha_{H}  = \frac{G_H}{2} \left( -\frac{\qc}{m} \, - \, 
\,  \frac{m^2}{4 \pi} \, + \, \frac{3(1-g_A)}{2 \rho} f_\pi^2
 \, - \, \frac{(3 \pi-8)) }{192 m^2} \gc
                           \right) \;. 
\end{equation}
In the limit $\Delta \rightarrow m $
\begin{equation}
\alpha_H \equiv  -  2iG_HN_c\left(  -  I_1  + 
 \fr{i }{96 N_c \, m^2}\gc\right) \; ,
\end{equation}
or, eliminating divergent integrals,
\begin{equation}
\alpha_{H}  = \frac{G_H}{2} \left( -\frac{\qc}{m} - \frac{\gc}{24 m^2}
                           \right) \;  . 
\end{equation}

This completes the specification and bosonisation of the HL$\chi$QM.
The remaining free parameters of the model are two 
condensates $\qc$ and $\gc$ and two constituent masses
$m$ and $\Delta$. We can
now apply the model to calculation of phenomenological quantities,
starting with the Isgur-Wise function.

\section{The Isgur-Wise function}
\label{sect:iw}

\begin{figure}[tb]
\centerline{\includegraphics[scale=0.6]{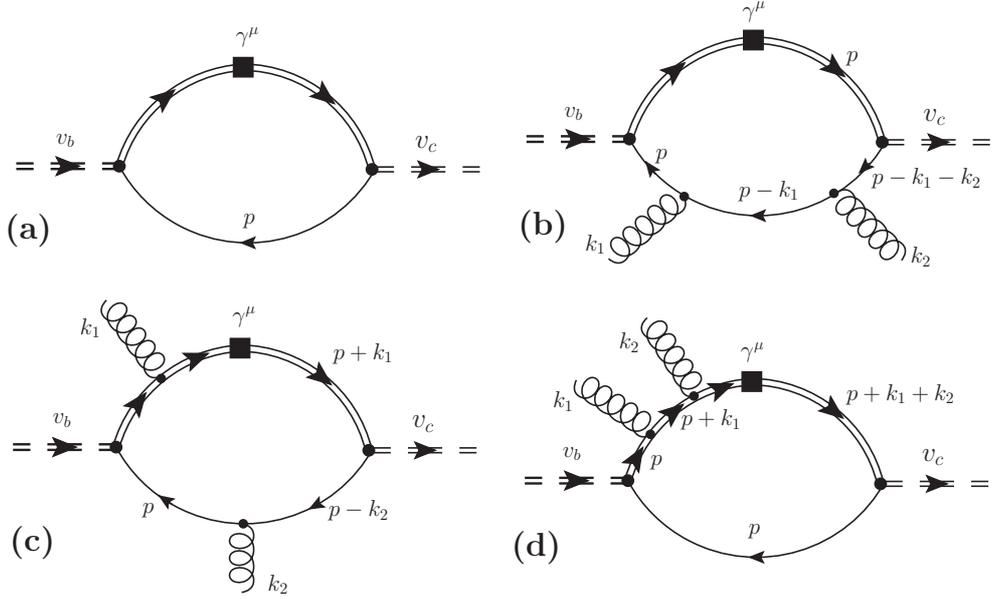}}
\caption{
Bosonisation 
corresponding to the Isgur-Wise function.
}
\label{fig:iwdiagrams}
\end{figure}

The Isgur-Wise function \cite{iw},
$\xi (\omega)$, relates all 
the form factors describing the processes $B\to D(D^*)$ in the
heavy quark limit.
In our framework it can be defined via bosonisation of
heavy-heavy quark current responsible for 
$B \rightarrow D$ transition:
\begin{equation}
 \overline{Q_{v_b}^{(+)}} \,\gamma^\mu\, L Q_{v_c}^{(+)}\;  \longrightarrow \;
    - \xi(\omega) Tr\left[
 \overline{H_c^{(+)}} \gamma^\mu L  H_{b}^{(+)} \right] \; \equiv \;
 J_{b \rightarrow c}^\mu \; \; .
\label{Jbc}
\end{equation}
Here $Q_{v_c}$ and $Q_{v_c}$ are the $c$ and $b$ quark fields within
HQEFT.
The IW function can be determined by calculating the
diagrams shown in figure (\ref{fig:iwdiagrams}).

One normally expects that emission of the soft gluons
from a heavy quark doesn't occur at the zeroth order in $1/m_Q$.
Namely, using
the HQET Lagrangian (\ref{LHQET}),
 and differentiating the expressions involving  heavy quark propagators
 according (\ref{eq:Afield}), will naturally,
for diagrams of same class as  those on Fig.~\ref{fig:f+}, lead to expressions
proportional to
$v_\mu v_\nu G^{a}_{\mu\nu}= 0$, where $v_\mu$ is either $v_b$ or $v_c$.
However, for the Isgur-Wise function there are two velocities
($v_b$ and $v_c$) in play in the diagram.
 Therefore one may obtain contributions
proportional to
\begin{equation}
 v_b^\mu v_c^\nu G^a_{\mu\nu} \; \; ,
\end{equation}
which will generally, away from the 
strict heavy quark limit  $\omega \rightarrow 1$, be different from zero.
Let us also mention that some care is required because
momentum flow in the diagrams for the translationally non-invariant
amplitudes in the Fock-Schwinger gauge is non-trivial,
see Fig.~\ref{fig:iwdiagrams}.

The corresponding results for diagrams (a)--(d) are
\begin{align}
\xi(\omega)_a & = i G_{H}^{2} N_c \left(
(m-\frac{2}{\omega+1} \Delta ) 
I_{1,1,1}-\frac{2}{\omega+1} I_{1,1}\right) \; \, ,\\
\xi(\omega)_b & = i G_{H}^{2} \gc m \pi^2 
\left( m^2I_{4,1,1} + I_{3,1,1}
- \frac{2 m}{\omega+1}\left(I_{4,1}+\Delta  I_{4,1,1}\right) \right) \; \, ,\\
\xi(\omega)_c & = \frac{i}{12} G_{H}^{2}
\gc  m \pi^2 (\omega -1) I_{2,2,2} \; \, , \\[1ex]
\xi(\omega)_d & = \frac{i}{12} G_{H}^{2}
\GluonCond \pi^2 (\omega -1) \bigg(-I_{1,2,3}
-I_{1,3,2}+\big(m (\omega+1) -2 \Delta \big) I_{1,3,3}\bigg) \;  \, .
\end{align}
Since loop integrals have at the least two heavy and one
light quark propagator, integrals of the type $I_{n,r,s}$
occur above, and they are defined in the 
Appendix~\ref{app:loop}.
We find that the identity $\xi(\omega = 1)=1$ follows from normalization of the
vector current in eq. (\ref{KX}):
\bea
 \xi(\omega=1) \, = \; \left( \xi_a  \, + \, \xi_b
 + \xi_c \, + \, \xi_d \right)_{\omega=1} \; = \;  \left( \xi_a
 \, + \,  \xi_b \right)_{\omega=1} \; =  \, 1 \; . 
\label{xi1}
\eea

Finally, for the slope of the IW function in the no-recoil limit
 $\omega\to 1$, we have 
\begin{align}
\xi'(1)_a & = i G_{H}^{2} N_c \left(
\frac{1}{2} I_{1,1} - \big(\frac{1}{3}m - \frac{5}{6}\Delta\big)I_{1,2}
-\frac{2}{3}\Delta(m-\Delta)\big(I_{1,3} - 2 m^2 I_{3,1}\big) 
\right) \; \, , \\
\xi'(1)_b & = i G_{H}^{2}  \gc m \pi^2 \left( 
\frac{1}{4}I_{2,4} + \frac{m}{6}(m-\Delta)I_{3,4} + \frac{m}{2}
\big( I_{4,1} + \Delta I_{4,2} \big)
\right) \; \, ,\\
\xi'(1)_c & = i G_{H}^{2}
\gc  \frac{m \pi^2}{12}  I_{2,4} \; \, \\[1ex]
\xi'(1)_d & = - i G_{H}^{2}
\gc \frac{\pi^2}{6}  \bigg(I_{1,5} - (m  - \Delta) I_{1,6}\bigg) \; \, .
\end{align}
All the $I_{n,r}$ integrals above can be evaluated using formulas from
Appendix \ref{app:loop}.

\section{Results and discussion}

\begin{figure}[tb]
\begin{center}
\includegraphics[scale=0.7]{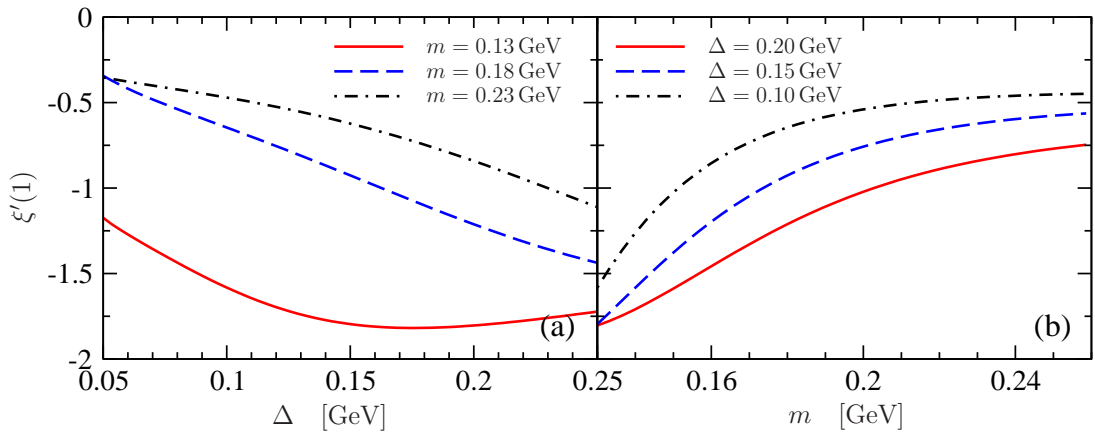}
\end{center}
\caption{The slope of Isgur-Wise function $\xi'(1)$ at no recoil point as a
function of $\Delta$ for three values of $m$ (a), and as a function of
$m$ for three values of $\Delta$ (b). The condensates are taken to be
$\qc = (-0.27\,{\rm GeV})^3$ and $\gc = (0.32\,{\rm GeV})^4$.
}
\label{fig:slope}
\end{figure}

Numerical results are presented in Figs. \ref{fig:cond} and \ref{fig:slope}.
Fig. \ref{fig:cond} displays the slope $\xi'(1)$ of IW function as a function
of the gluon condensate and the heavy meson decay constant $f_B$ as a function of 
the quark condensate, while Fig. \ref{fig:slope} shows slope $\xi'(1)$ 
for some generic condensate values in dependence
on the dynamical masses $\Delta$ and $m$. One observes that for reasonable
intervals of masses\footnote{Constituent mass of quarks in the presented
model is smaller than in other similar quark models due to the explicit 
inclusion of gluon condensate in the dynamics of the model.}
slope is of the order of -1 and beyond, which is
in agreement with values mentioned in the Introduction, as well as
with values obtained within other theoretical frameworks 
\cite{Morenas:1997nk} and on the lattice \cite{Bowler:2002zh}.
It should be noticed that values for $\xi'(1)$ of order or bigger than 
one prefer smaller constituent mass than in \cite{AHJOE}.
For easier overview of analytical results, as well as for their numerical checks,
it is convenient to again investigate limit $\Delta\to m$.
After using constant values for integrals in this limit, as
given in Appendix~\ref{app:loop} one obtains simple expression

\begin{equation}
\xi'(1)  =  \frac{(3 g_{{\cal A}} -1)}{4} - 
\frac{(1 + 3 g_{{\cal A}})}{4 f_{\pi}^2}
\left\{ \frac{m^2 N_c}{2 \pi^2} + \frac{11}{315 m^2} \gc
\right\} \;.
\end{equation}
Numerical values in this limit are not unreasonable.

In conclusion, we have presented an improved Heavy-Light Chiral Quark 
Model where
introducing additional mass parameter in the heavy-quark propagator
resulted in a flexible model capable of consistent description of
heavy-meson decays, where we placed particular emphasis on a 
characterization of Isgur-Wise function. Further applications of
the model, such as calculation of non-factorizable amplitudes for non-leptonic
heavy-meson decays could now be attempted.
As the slope of the IW function is steeper than the one used in,
say, Ref. \cite{EFP}, the partial amplitudes for $B \rightarrow D \overline{D}$ 
depending on the IW function, might be overestimated there. This will then have
 consequences for the size of the overall amplitude.

\subsection*{Acknowledgement}
K.K.~is grateful for a warm hospitality of the Department of Physics at 
the University of Oslo.
This work was supported by Research Council of Norway  
and the Croatian Ministry of
Science, Education and Sport, contract no. 119-0982930-1016.

\appendix 
\apndxeqn

\section{Loop integrals}
\label{app:loop}

Three-, two-, and one-point loop integrals with one light-quark
propagator  occurring in the calculations are:
\begin{align}
I_{n,r,s}^{\alpha \beta \cdots}&  =
\int \frac{\textrm{d}^4 p}{(2 \pi)^4}\, \frac{p^{\alpha} p^{\beta} \cdots}
{(p^2 - m^2 +i\epsilon)^n\, (p\cdot v - \Delta +i\epsilon)^r\, 
(p\cdot v' - \Delta +i\epsilon)^s} \;, \\
I_{n,r}^{\alpha \beta \cdots}&  =
\int \frac{\textrm{d}^4 p}{(2 \pi)^4}\, \frac{p^{\alpha} p^{\beta} \cdots}
{(p^2 - m^2 +i\epsilon)^n\, (p\cdot v - \Delta +i\epsilon)^r} \;, \\
I_{n}^{\alpha \beta \cdots}&  =
\int \frac{\textrm{d}^4 p}{(2 \pi)^4}\, \frac{p^{\alpha} p^{\beta} \cdots}
{(p^2 - m^2 +i\epsilon)^n} \;.
\label{eq:intdel}
\end{align}
To evaluate such integrals
one first reduces tensor to scalar ones using relations ($\omega\equiv v\cdot v'$):
\begin{align}
I^{\alpha}_{n, s}& = (I_{n, s-1} + \Delta I_{n, s}) v^{\alpha} \;  \, , \\
I^{\alpha\beta}_{n, 1}&  =  
\frac{1}{3} \Big[I_{n-1,1} + m^2 I_{n,1} - 
\Delta I_n - \Delta^2 I_{n,1} \Big] g^{\alpha \beta } \; \, , \nonumber \\
 & \quad - \frac{1}{3} \Big[I_{n-1,1}+m^2 I_{n,1}-4 \Delta I_n
  -4 \Delta^2 I_{n,1} \Big] v^{\alpha} v^{\beta} \; \, , \\
I^{\alpha\beta\gamma}_{n, 1}&  = 
\frac{1}{2} \Big[-I_{n-1}-(m^2-4\Delta^2)I_n - 2\Delta I_{n-1,1}
- 2\Delta(m^2-2\Delta^2)I_{n,1}\Big] v^\alpha v^\beta v^\gamma \; \, ,
  \nonumber \\
& \quad + \frac{1}{3} \Big[ \frac{3}{4} I_{n-1} + 
\big(\frac{3}{4}m^2 - \Delta^2\big) I_n + \Delta I_{n-1,1} \nonumber \\
& \qquad \qquad \qquad +\Delta(m^2 -\Delta^2)I_{n,1}\Big]
(v^\alpha g^{\beta\gamma} + v^\beta g^{\alpha\gamma} +
v^\gamma g^{\alpha\beta}) \\
I^{\alpha}_{n,r,s}& = \frac{1}{\omega^2 - 1} 
 \bigg\{
 (\omega I_{n,r,s-1}-I_{n,r-1,s})v^\alpha +
  (\omega I_{n,r-1,s}-I_{n,r,s-1})v'^\alpha \; \, , \nonumber \\
 & \qquad \qquad \qquad + 
\Delta(\omega-1)I_{n,r,s}(v^\alpha + v'^\alpha) \bigg\} \; \, .
\label{eq:tensorred}
\end{align}
(Reduction of one-point $I_{n}^{\alpha \beta \cdots}$ tensor
integrals is simple and well-known.)
Now all scalar two-point integrals $I_{n,s}$ can be
reduced to linear combinations of $I_{n,1}$ integrals using
general recursion formula
\begin{multline}
I_{n,s} = \frac{-4 n}{s^2 -3 s +2} \bigg\{(n+s-3)
I_{n+1,s-2} + m^2(n+1) I_{n+2,s-2}  \\
+\Delta \frac{s^2 - 3s
+2}{s-1} I_{n+1,s-1} \bigg\} \;,
\end{multline}
valid for $s>2$, whereas the special case $s=2$ is
\begin{equation}
I_{n,2} = -2 n \big( I_{n+1} + \Delta I_{n+1,1} \big)  \;. 
\end{equation}

For the calculation of the slope $\xi'(1)$ of Isgur-Wise function, 
one additionally  needs derivatives of three-point integrals
at point $\omega=1$. Integrals themselves at this point are trivially 
given by 
\begin{equation}
I_{n,r,s}(1) \; =   \; I_{n,r+s}
\end{equation}
 The derivatives are given by
\begin{equation}
\left.\frac{\partial I_{n,r,s}(\omega)}{\partial \omega}\right|_{\omega=1} =
\frac{r s}{2 (n-1)} I_{n-1,r+s+2} \qquad {\rm for} \quad n>1 \;,
\end{equation}
while for $n=1$ we have special case:
\begin{equation}
\left.\frac{\partial I_{1,r,s}(\omega)}{\partial \omega}\right|_{\omega=1} =
  - \frac{r s}{r+s+1}  \big(2\frac{r+s-1+\epsilon}{r+s} I_{1,r+s} 
+ 2\Delta I_{1,r+s+1} + \frac{2 m^2}{r+s} I_{2,r+s} \big) \;,
\end{equation}
where dimensional regularization parameter $\epsilon$ is relevant only
for divergent case $r=s=1$ where using 
$\lim_{\epsilon\to 0}\epsilon I_{1,2}=-i/(8\pi)$
one gets
\begin{equation}
\left.\frac{\partial I_{1,1,1}(\omega)}{\partial \omega}\right|_{\omega=1} =
  - \frac{1}{3}  \big( I_{1,2} + 2\Delta I_{1,3} - 4 \Delta m^2 
I_{3,1} \big) \;.
\end{equation}
Then again recursion formulas above above can be used to reduce 
everything to $I_{n,1}$ integrals. These integrals can be explicitly
evaluated and they read
\begin{align}
I_{n,1}& \equiv \frac{i}{16 \pi^2 m^{2 n-3}}\: 
a_{n}(\frac{\Delta}{m}) \;, \qquad \text{where} \\[1.5ex]
a_2(x) & = - \frac{1}{1-x^2} \mathcal{F}(x) \quad\xrightarrow{x\to 1} -2 \\
a_3(x) & = \frac{1}{4} \frac{1}{(1-x^2)^{2}}                
\left( \mathcal{F}(x) - 2 x +2 x^3
\right) \quad\xrightarrow{x\to 1} \frac{1}{3}\\
a_4(x) & = -\frac{1}{24} \frac{1}{(1-x^2)^{3}}                
\left( 3 \mathcal{F}(x) - 10 x 
+14 x^3 - 4 x^5 \right)  \quad\xrightarrow{x\to 1} -\frac{2}{15}\\
a_5(x) & = \frac{1}{192} \frac{1}{(1-x^2)^{4}}                
\left( 15 \mathcal{F}(x) - 66 x +118 x^3 - 68 x^5 + 16 x^7
\right)  \quad\xrightarrow{x\to 1} \frac{1}{14} \\
a_6(x) & = - \frac{1}{1920} \frac{1}{(1-x^2)^{5}}
\left( 105 \mathcal{F}(x) - 558 x 
+1210 x^3 \right. \nonumber \\
& \qquad\qquad\qquad\qquad\qquad
 \left. - 1052 x^5
+ 496 x^7 -   96 x^9 \right)  \quad\xrightarrow{x\to 1} -\frac{2}{45} \;.
\end{align}

Here the function $\mathcal{F}(x)$ is \cite{Bouzas:1999ug}
\begin{align}
\mathcal{F}(x)& = \sqrt{x^2 - 1 + i\epsilon}\Big(
\log(x - \sqrt{x^2 -1 +i\epsilon}) - \log(x + \sqrt{x^2-1+i\epsilon})
\Big) 
\end{align}
Note that $\mathcal{F}(x)\xrightarrow{x\to 0} \pi $, which gives the 
various $a_n(0)$. Furthermore, 
          $\mathcal{F}(x) = - 2 x F(1/x)$, where $F(1/x)$ is
function from Eq. (A2) of \cite{Stewart:1998ke}.

This reduction of integrals is easy to implement on a computer and
corresponding Mathematica code is available from the authors upon
request.

\bibliographystyle{unsrt}

\end{document}